\documentclass[conference]{IEEEtran}
\IEEEoverridecommandlockouts

\usepackage{listings}

\usepackage{cite}
\usepackage{amsmath,amssymb,amsfonts}
\usepackage{algorithmic}
\usepackage{graphicx}
\usepackage{textcomp}
\usepackage{xcolor}

\usepackage{array}
\usepackage{enumitem}
\usepackage{multirow}
\usepackage{caption}
\usepackage{booktabs}
\usepackage{pgfplots}
\pgfplotsset{compat=1.18}
\usepackage{MnSymbol}

\usepackage{hyperref}
\hypersetup{colorlinks=false}
\hypersetup{linkcolor=red}
\hypersetup{urlcolor=red}
\hypersetup{citecolor=red}

\def\BibTeX{{\rm B\kern-.05em{\sc i\kern-.025em b}\kern-.08em
    T\kern-.1667em\lower.7ex\hbox{E}\kern-.125emX}}
\begin{document}

\title{RISC-V processor enhanced with a dynamic micro-decoder unit}

\author{\IEEEauthorblockN{Juliette POTTIER\dag, Thomas NIEDDU\ddag, Bertrand LE~GAL\ddag, Sébastien PILLEMENT\dag, and Maria MENDEZ REAL\dag}
\IEEEauthorblockA{\dag Nantes Université, CNRS, IETR-UMR 6164, F-44000 Nantes, France\\
\ddag Université de Rennes, IRISA/INRIA lab., CNRS UMR 6074, 35042 Rennes, France}
}

\maketitle

\begin{abstract}
For years, the open-source RISC-V instruction set has been driving innovation in processor design, spanning from high-end cores to low-cost or low-power cores. After a decade of evolution, RISC architectures are now as mature as the CISC architectures popularized by industry giant Intel. Security and energy efficiency are now joining execution speed among the design constraints. In this article, we assess the benefits and costs associated with integrating a micro-decoding unit inspired by CISC processors into a RISC-V core. This unit, added in a specific pipeline stage, should enable dynamic custom instruction sequences execution whose usage could be, for instance to compress binaries, obfuscate behavior, etc.
\end{abstract}

\begin{IEEEkeywords}
Architecture, RISC-V, ISA, CISC, micro-decoding
\end{IEEEkeywords}

\section{Introduction}

The field of Internet of Things (IoT) \cite{hasan_2022} and Edge Computing \cite{edge} requires the production of a wide range of processor architectures that meet various application constraints (computing power, cost, energy, security, etc.). Currently, Complex Instruction Set Architecture (CISC) and Reduced Instruction Set Architecture (RISC) processors provide high levels of performance, but in two different ways \cite{Burr04}. The efficiency of CISC architectures comes from their ability to implement complex operations with a single assembler instruction. They provide the compiler with many macro-instructions that are efficiently decoded and executed by the hardware architecture. These macro-instructions offer advantages from an application perspective; for example, they allow for optimal utilization of the hardware architecture on which the application is executed, and they also enable a high density of binary programs. These architectures are mainly developed by large industrial groups such as Intel or AMD and are therefore closed. Efforts from the scientific community have enabled the extraction of microcode from AMD K8 and K10 processors and understanding the operation of their micro-decoding unit \cite{reverse_x86}. While being transparent to the end user, the microcode (stored in dedicated a memory) can be updated by the manufacturer to fix potential bugs and/or counter certain post-fabrication attacks. This is notably what allowed Intel to address the security vulnerabilities SPECTRE \cite{spectre} and MELTDOWN \cite{meltdown}.

RISC architectures are designed to integrate few instructions, thus lacking the micro-decoding mechanism. The open-source RISC-V ISA (Instruction Set Architecture), in its RVI configuration, provides the compiler with about fifty elementary instructions \cite{Harr21}. Many architectures, both low-cost~\cite{lowcost_rv4, lowcost_rv, lowcost_rv3, lowcost_rv2, trojan} and high-end~\cite{cva6, boom, blackparrot}, implement this instruction set. Currently, high-end RISC-V processors feature 64-bit data paths, deep pipelines, and are capable of running a Linux-type operating system thanks to their advanced architectural optimizations. When using all these architectures, it is up to the compiler to identify the appropriate instruction combinations to generate efficient code. This inevitably leads to the production of larger programs compared to their CISC counterparts, despite the use of compressed instruction formats~\cite{code:comp,density}. Furthermore, unlike CISC architectures, hardware implementations of this reduced instruction set do not allow for post-fabrication architecture updates necessary, for example, to address security vulnerabilities, fix bugs, better adapt to the application domains or support new instruction set extensions.

In this paper, we investigate the introduction of a micro-decoder unit into RISC-V processor architectures to provide flexibility. This mechanism, introduced into the architecture, enables the decoding of custom instructions to RISC-V instructions, or from RISC-V instructions to RISC-V instructions. This work would allow the designer to: (1) dynamically alter the behavior of the micro-architecture without modifying the original binary, (2) compress the size of programs, or (3) modify the execution of an instruction. To our knowledge, there is no existing work in the literature addressing the integration of such a unit into a RISC processor.

The rest of the article is structured as follows. Section~\ref{sec:Context} presents the study's context and the selected RISC-V core architecture. Section~\ref{sec:design} discusses the strategies for integrating the micro-decode unit, our chosen approach, and the necessary architectural modifications. The testing environment, experimental methodology, and initial results are detailed in Section~\ref{sec:Results}. Section~\ref{sec:Conclusion} discusses these results before concluding.

\section{Micro-decoding unit proposal}
\label{sec:Context}

\begin{figure*}[!tb]
    \centering
    \includegraphics[width=.9\textwidth]{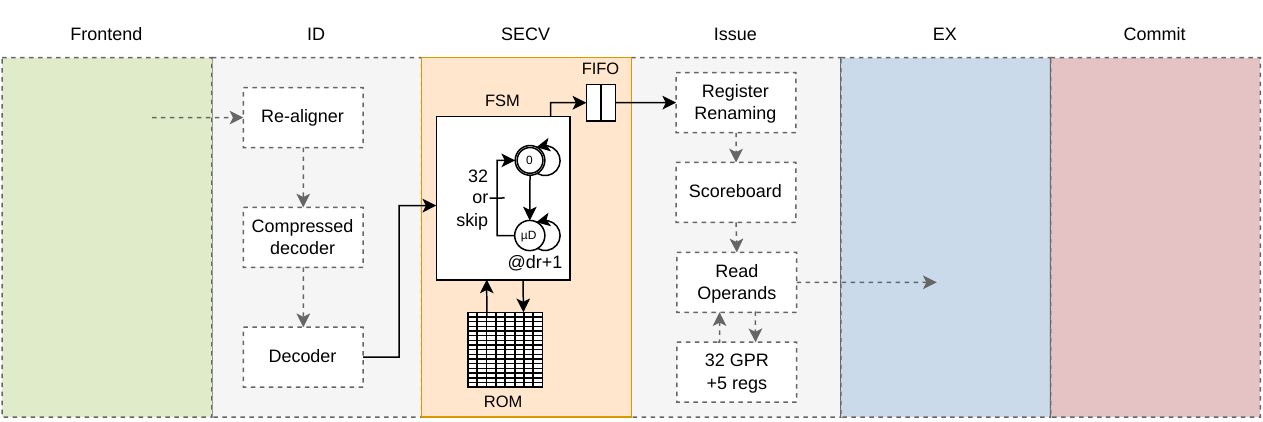}
    \caption{CV64A6 core architecture enhanced with our micro-decoding unit}
    \label{fig:pipeline_secv}
\end{figure*}

Various high-end processor architectures compatible with the RISC-V instruction set are currently available. The most popular processor cores are BOOM \cite{boom}, CVA6 \cite{cva6}, SHAKTI and ROCKET cores. To achieve high-performance levels, they incorporate many pipeline stages that partition the computations to be executed. The number of pipeline stages generally varies from 5 to a dozen, depending on the core setup. In the most traditional configuration, architectures supporting compressed instructions, such as the CVA6 \cite{cva6}, feature 6 \textit{pipeline} stages. The first 2 stages, called \textit{Frontend}, handle PC generation, instruction cache access, branch prediction and an instruction queue. The following stage (\textit{ID}) realigns compressed instructions and translates instructions into control signals. The next stage, Issue, integrates the \textit{scoreboard}. The \textit{scoreboard} controls the next 2 stages, scheduling the operations to be executed according to available resources. This is where the register queues (GPR and FPR) are accessed. The execution stage (\textit{EX}) takes care of the execution of the operations by the functional units. Finally, the last stage, \textit{Commit}, removes the instructions from the \textit{pipeline} and retrieves the pending results.

The addition of a CISC-like micro-decoding unit can be introduced at different levels. The simplest strategy is to integrate it ahead of the instruction decoder (after the fetch stage). The main advantage of this approach is that macro-instructions can be easily detected, as they are still compressed to 32 bits. In addition, it is possible to reuse the traditional instruction decoder to generate the control signals that will drive the architecture. However, this approach has some notable drawbacks. The main limitation is the impossibility of addressing additional registers outside GPR space to store temporary computation data, thus limiting the interest of the approach. Indeed, it discards the possibility to store temporary values. Consequently, it was decided to add a micro-decoding unit between the Decode stage and the Issue one, as shown in Figure \ref{fig:pipeline_secv}. A specific pipeline stage, dedicated to this task, is added to provide both (1) flexibility and modularity and (2) to avoid impacting the critical path of the core.

\section{Microdecoding unit design} \label{sec:design}

The micro-decoding unit must be able to generate sequences of $\mathbf{N_P}$ instructions for each of the $\mathbf{P}$ macro-instructions it needs to handle. Different strategies are possible for generating such sequences, such as the use of hardwired logic. However, hardware complexity will quickly increase with the values of $\mathbf{N_P}$ and $\mathbf{P}$, and the architecture of the micro-decoding stage will need to be revised with each modification of the sequences. Moreover, this approach disables post-design microcode upgrade. The most generic approach is based on the use of memory, which is responsible for storing the $\mathbf{P}$ sequences of $\mathbf{N_P}$ micro-instructions.

\begin{figure*}
    \centering
    \includegraphics[width=.9\textwidth]{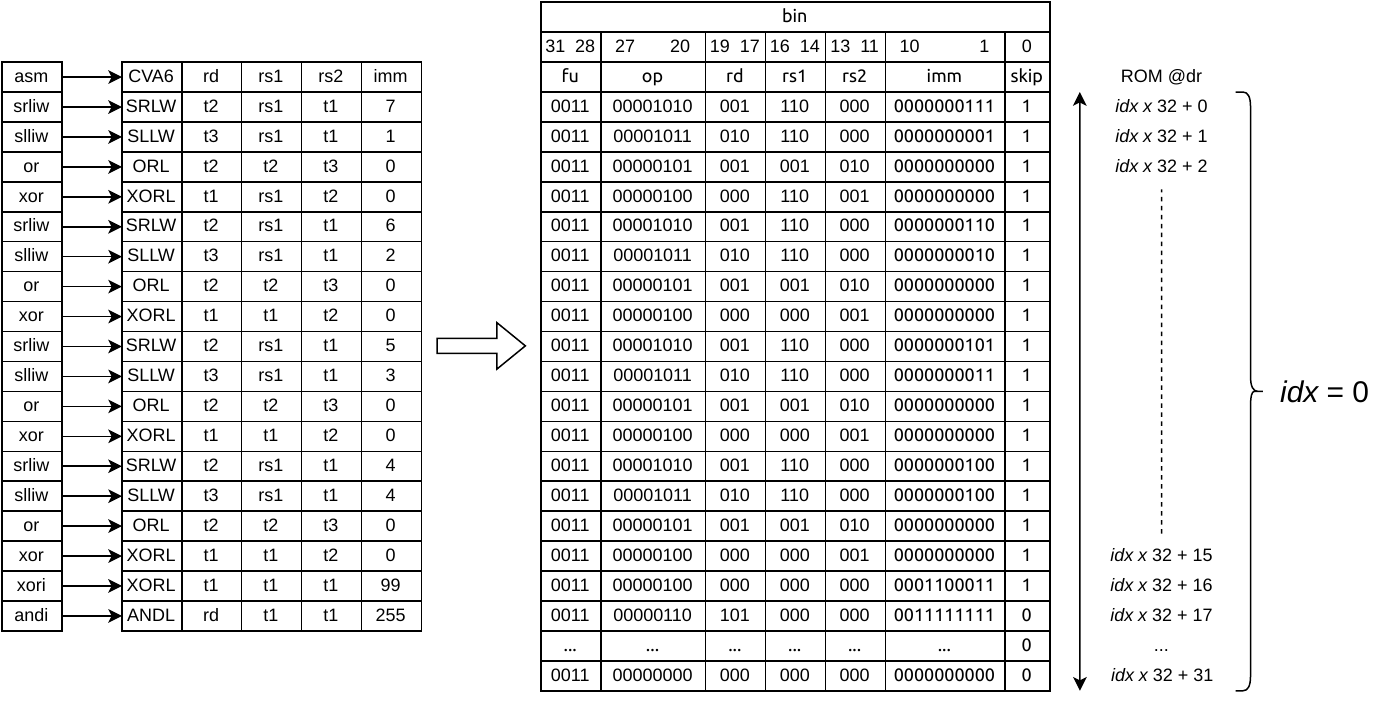}
    \caption{Micro-decoding unit ROM content for macro-instruction encoding for S-Box computation}
    \label{fig:rom}
\end{figure*}

Assuming that each macro-instruction is made up of a maximum of $\mathbf{N_P}$ micro-instructions, and that for address simplicity, macro-instructions reserve $\mathbf{P}$ slots for all macro-instructions, then memory is made up of $\mathbf{N_P} \times \mathbf{P}$ elements. 
Instructions relating to the macro-instruction $\mathbf{idx}$ are stored in the address range $[ \mathbf{idx} \times \mathbf{N}, (\mathbf{idx}+1) \times \mathbf{N}-1 ]$. Each binary word in this memory represents an instruction in the sequence $\mathbf{N_P}$. To reduce the hardware complexity of the microdecoding unit, it is important that these binary words are as compact as possible. However, they must first indicate the functional unit on which the operation is to be performed (e.g. \emph{ALU} (\textit{Arithmetic-Logic Unit}), \emph{LSU} (\emph{Load-Store Unit}), \ldots) and the operation to execute (e.g. ADD, XORI, \ldots). In the case of the CVA core, this information is encoded using the \emph{fu} and \emph{opcode} fields, as shown in Figure \ref{fig:rom}. Initially, we assume that the sequences of arithmetic and logical operations for the two fields can be stored on 4 bits and 8 bits, respectively.

In addition, it is necessary to memorize the source (\texttt{rs1} and \texttt{rs2}) and destination (\texttt{rd}) registers to be used by the micro-instructions. 
When microdecoding a macro-instruction, the ${rd}$ register provisioned by the compiler for macro-instruction execution may not be enough to store temporary data. 
Consequently, the addresses of the \texttt{rd}, \texttt{rs1} and \texttt{rs2} registers are encoded on 3 bits, enabling each instruction to access \texttt{rs1}, \texttt{rs2}, \texttt{rd} as well as 5 additional registers (added to the GPR) which are only accessible via the microdecoder. Finally, to make the microdecoder more flexible, two additional fields have been added. The first (\texttt{imm}), allows manipulation of immediate values, while the second (\texttt{skip)} indicates the current instruction is the last valid instruction in the sequence.

Without further optimization of the microdecoder's memory structure, and assuming that immediate data is encoded on $10$ bits, the $\mathbf{M}$ size of each microinstruction is $32$ bits. The overall cost of memory would evolve linearly with $\mathbf{P}$: $\mathbf{P} \times \mathbf{N} \times \mathbf{M} = \mathbf{P} \times \mathbf{N} \times 32$ bits. This format is consistent with the size of the memory blocks available on FPGA circuits. It should be noted that it would be possible to implement special coding of micro-instructions (type-R, type-I, etc.) to minimize memory complexity, but this would require decoding logic.

The microinstruction memory is controlled and interfaced with the processor pipeline by a state machine (FSM: Finite State Machine). This machine has two modes of operation: (1) a \textit{bypass} mode that transmits conventional RISC-V instructions to the next \textit{pipeline} stage; and (2) a mode for injecting micro-instructions into the \textit{pipeline}, which is triggered as soon as a macro-instruction is detected.     The microdecoding mode includes address calculation based on the sequence to be injected. Synchronization with other stages can be achieved using a register barrier or even a register queue of the \textit{FIFO} (\textit{First In First Out}) type.

 As depicted in figure \ref{fig:pipeline_secv}, the additional \textit{pipeline} stage integrates 3 elements:
 
\begin{enumerate}[leftmargin=4mm]
    \item a \textit{FIFO} in charge of synchronization (acknowledgement signals) with the \textit{Issue} stage. It enables instructions to be delayed, as the microdecoder can generate one instruction per cycle, thus saturating the \textit{scoreboard}. Its depth has been set at 2, while its width is linked to the control signals carried in the \textit{pipeline}, i.e. 364 bits. This is a \textit{FIFO} of the \textit{FWFT} type (\textit{First-Word-Fall-Through}), minimizing the data access penalty.

    \item an FSM associated with an address counter. In its initial state, it behaves like a register barrier: it stores the relevant information (PC, immediate value, \texttt{rd}, \texttt{rs1}, \texttt{rs2}, \ldots) of decoded instructions. When a decoded macro-instruction is presented, it triggers a microdecoding sequence. Based on its $\mathbf{idx} \in \mathbf{P}$, the FSM calculates the address corresponding to the correct sequence in memory and enters the microdecoding state. When the sequence run is complete, the FSM returns to the (\textit{bypass}) state.

    \item the memory containing the sequences of micro-instructions described in the previous section. An example of memory from one of the test applications (AES encryption/decryption) is shown in Figure \ref{fig:rom}. This calculation generates the contents of the S-Box. It requires logical operations: SHIFT, AND, OR, XOR in order to rotate on 8 bits ($\rcirclearrowleft$). This operation is broken down into 18 RISC-V instructions. In the sequence, the registers marked $t$ refer to temporary registers accessible only via the microdecoding unit.
\end{enumerate}

%
\section{Experimental results}
\label{sec:Results}
%
The impact of inserting the micro-decoding unit inside a new \textit{pipeline} stage was evaluated in terms of execution time penalties of a set of test applications. To achieve this, we modified a CV64A6 \cite{cva6} core. The architecture was simulated at cycle level using the Verilator tool (version 4.110) to accurately evaluate the execution times of various benchmarks from the literature \cite{papierBertrand, benchs:Mi, benchs:trap, lin2004error}. Benchmark execution times were calculated from CSR registers. The \textit{toolchain} used to generate the test programs is based on GCC compiler version 12.1.0.

The results of the  cycle accurate simulations show that when the micro-decoding unit is included in the CVA6 core, it has a slight impact on program execution time, even if the program does not use the micro-decoding unit. A slight variation of up to +0.3\% was observed on 25 benchmarked applications, This penalty is due to the time taken to re-fill the pipeline in the event of incorrect branch prediction at runtime. The same protocol was followed for the two pedagogical examples, including the S-Box generation and an AES encryption using the microdecoding unit. In simulation, without using the microdecoder, the blank CV64A6 completes the execution in 55368 clock cycles. On the proposed architecture, when the micro-decoding unit is used to deliver the 18-instruction sequence, the timing is 54796 clock cycles. Performance is better (-1\% less clock cycles) when it is used. This is due to the fact that ROM injects one instruction per clock cycle into the \textit{pipeline}, whereas the blank architecture has to fetch them from its instruction cache or even its main memory, incurring penalties. This finding is similar when the study is carried out directly on an FPGA board. However, unlike simulation results \textit{cycle-accurate}, those obtained on circuit \textit{FPGA} are more difficult to obtain and interpret. This is due to the OS, and more particularly the scheduler, which due to its regular execution noises the measurements, making it impossible to obtain accurate data when DDR memory and caches are no longer simulated.

The impact on the core's hardware complexity was also investigated. The architecture was synthesized and implemented on a Kintex-7 FPGA (XC7K325T-2FFG900C) using Xilinx's Vivado 2020.2 tool. A summary of the results obtained after placement and routing is provided in Table \ref{tab:hardCore2}. In this assessment, the maximum operating frequency set at 50 MHz is not impacted by the micro-decoding stage, demonstrating the absence of a critical path in our stage.

\begin{table}
    \centering
 \begin{tabular}{@{}rrrrrrr@{}}
        & LUT & SRL & FF & BRAM36 & BRAM18 & DSP\\
        \toprule
            baseline
            & 44148 &     0 & 24076     & 36 & 0 & 27 \\
        \midrule
            With $\mu$dec.
            & 45880   &     0 &   25078 &    36 &  0 & 27 \\
            (2 instr.)& \textcolor{red}{+3,9\%}  & -     &  \textcolor{red}{+4,2\%} &     - &  - &  - \\
        \midrule
            With $\mu$dec.
           &  45922 & 0 & 25046   &     37 &  0 & 27 \\
           (32 instr.) & \textcolor{red}{+3,8\%} & -  & \textcolor{red}{+3,8\%} &    \textcolor{red}{+2,6\%} &  - &   - \\
        \midrule
         With $\mu$dec.
           &  45937 & 0 & 25068   &     38 &  0 & 27 \\
           (64 instr.) & \textcolor{red}{+4,1\%} & -  & \textcolor{red}{+4,1\%} &    \textcolor{red}{+5,6\%} &  - &   - \\
        \bottomrule
    \end{tabular}
    \caption{Hardware complexity comparison}
    \label{tab:hardCore2}
\end{table}  

The first experiment illustrates the case where the micro-decoding unit has only 2 micro-instructions. In this context, there was a slight increase in hardware complexity of 3.9\% for LUTs and 4.2\% for FFs. These increases are mainly due to the addition of the FSM, 5 registers in the GPR queue (and its multiplexing logic) and the insertion of a \textit{FIFO} of depth 2 and width 364 bits between the microdecoding stage and the execution stage. Note that in its version with more than 2 micro-instructions, the memory of our stage is implemented by distributed memory. The type of implementation changes as the size of the memory containing the micro-instructions increases. For example, when 64 macro-instructions are stored, 2 RAM blocks are added to the LUTs and FFs.

%
\section{Conclusion}\label{sec:Conclusion}
%
    
In this article, a CISC-type microdecoding unit has been implemented in a RISC-V ecosystem processor. This solution, involving the addition of an extra pipeline slice, results in only a small latency overhead for applications not using this functionality. Moreover, the additional costs in terms of hardware complexity are low in comparison with the complexity of the CVA6 core. This work shows that it is possible to hybridize a RISC-V architecture by adding a microdecoder. This enhancement opens the way to new research perspectives for: (1) binary size reduction, identifying recurrent instruction patterns and incorporating them into the micro-decoding memory (2) for static code obfuscation purposes, adding specialized instructions, akin to “black boxes,” (3) to address security concerns for instance by injecting phantom instructions \cite{bossuet, mem_frencing} into the processor pipeline and/or alter the way calculations are performed to complicate attacks via side-channels.

\section*{Acknowledgment}
We thank the ANR which supports this work under the convention ANR-21-CE-39-0017.

\bibliographystyle{IEEEtran}
\bibliography{riscv}

\begin{thebibliography}{10}
\providecommand{\url}[1]{#1}
\csname url@samestyle\endcsname
\providecommand{\newblock}{\relax}
\providecommand{\bibinfo}[2]{#2}
\providecommand{\BIBentrySTDinterwordspacing}{\spaceskip=0pt\relax}
\providecommand{\BIBentryALTinterwordstretchfactor}{4}
\providecommand{\BIBentryALTinterwordspacing}{\spaceskip=\fontdimen2\font plus
\BIBentryALTinterwordstretchfactor\fontdimen3\font minus \fontdimen4\font\relax}
\providecommand{\BIBforeignlanguage}[2]{{%
\expandafter\ifx\csname l@#1\endcsname\relax
\typeout{** WARNING: IEEEtran.bst: No hyphenation pattern has been}%
\typeout{** loaded for the language `#1'. Using the pattern for}%
\typeout{** the default language instead.}%
\else
\language=\csname l@#1\endcsname
\fi
#2}}
\providecommand{\BIBdecl}{\relax}
\BIBdecl

\bibitem{hasan_2022}
M.~Hasan, ``{State of {IOT} 2022: Number of connected {IOT} devices growing 18\% to 14.4 billion globally},'' https://iot-analytics.com/number-connected-iot-devices, 2022, consult{\'e}: 29 Mars 2023.

\bibitem{edge}
V.~N. Chander and K.~Varghese, ``A soft {RISC-V} vector processor for {Edge-AI},'' in \emph{Proceedings of VLSID}, 2022.

\bibitem{Burr04}
M.~Burrell, \emph{Fundamentals of Computer Architecture}.\hskip 1em plus 0.5em minus 0.4em\relax Red Globe Press London, 2004.

\bibitem{reverse_x86}
P.~Koppe, B.~Kollenda, and {al.}, ``{Reverse Engineering x86 Processor Microcode},'' \emph{CoRR}, vol. abs/1910.00948, 2019.

\bibitem{spectre}
P.~Kocher, J.~Horn, and {al.}, ``{Spectre Attacks: Exploiting Speculative Execution},'' in \emph{Proceedings of S\&P'19}, 2019.

\bibitem{meltdown}
M.~Lipp, M.~Schwarz, and {al.}, ``{Meltdown: Reading Kernel Memory from User Space},'' in \emph{Proceedings of USENIX Security}, 2018.

\bibitem{Harr21}
S.~L. Harris and D.~Harris, \emph{Digital Design and Computer Architecture, RISC-V Edition}.\hskip 1em plus 0.5em minus 0.4em\relax Elsevier Inc., 2021.

\bibitem{lowcost_rv4}
Y.-H. Cheng, L.-B. Huang, and {al.}, ``{RV16: An Ultra-Low-Cost Embedded RISC-V Processor Core},'' \emph{Journal of Computer Science and Technology}, vol.~37, no.~6, pp. 1307--1319, 2022.

\bibitem{lowcost_rv}
K.~Patsidis, D.~Konstantinou, and {al.}, ``{A low-cost synthesizable RISC-V dual-issue processor core leveraging the compressed Instruction Set Extension},'' \emph{Microprocess. Microsystems}, vol.~61, pp. 1--10, 2018.

\bibitem{lowcost_rv3}
D.~A. Santos, L.~M. Luza, and {al.}, ``{A Low-Cost Fault-Tolerant RISC-V Processor for Space Systems},'' in \emph{Proceedings of DTIS}, 2020.

\bibitem{lowcost_rv2}
R.~Serrano and {al.}, ``{A Low-Power Low-Area SoC based in RISC-V Processor for IoT Applications},'' in \emph{Proceedings of ISOCC}, 2021.

\bibitem{trojan}
N.~Albartus, C.~Nasenberg, and {al.}, ``On the design and misuse of microcoded (embedded) processors {\textemdash} a cautionary note,'' in \emph{Proceedings of USENIX Security}, 2021.

\bibitem{cva6}
F.~{Zaruba} and L.~{Benini}, ``{The Cost of Application-Class Processing: Energy and Performance Analysis of a Linux-Ready 1.7-GHz 64-Bit RISC-V Core in 22-nm FDSOI Technology},'' \emph{IEEE Transactions on Very Large Scale Integration (VLSI) Systems}, 2019.

\bibitem{boom}
J.~Zhao, B.~Korpan, and {al.}, ``{SonicBOOM: The 3rd Generation Berkeley Out-of-Order Machine},'' in \emph{Proceedings of the Workshop on Computer Architecture Research with RISC-V}, 2020.

\bibitem{blackparrot}
D.~Petrisko, F.~Gilani, M.~Wyse, and {al.}, ``{BlackParrot: An Agile Open-Source RISC-V Multicore for Accelerator SoCs},'' \emph{IEEE Micro}, vol.~40, no.~4, pp. 93--102, 2020.

\bibitem{code:comp}
H.~Lekatsas and W.~Wolf, ``Code compression for embedded systems,'' in \emph{Proceedings of DAC}, 1998.

\bibitem{density}
H.~Lozano and M.~Ito, ``Increasing the code density of embedded risc applications,'' in \emph{Proceedings of ISORC}, 2016.

\bibitem{papierBertrand}
\BIBentryALTinterwordspacing
B.~Le~Gal and C.~Jego, ``{Softcore Processor Optimization According to Real-Applicaion Requirements},'' \emph{IEEE Embedded Systems Letters}, vol.~5, no.~1, pp. 4--7, 2013. [Online]. Available: \url{https://hal.science/hal-00945635}
\BIBentrySTDinterwordspacing

\bibitem{benchs:Mi}
M.~Guthaus, J.~Ringenberg, and {al.}, ``{MiBench: A free, commercially representative embedded benchmark suite},'' in \emph{Proceedings of WWC-4 (Cat. No.01EX538)}, 2001.

\bibitem{benchs:trap}
{Fossati, Luca}, ``{TRAP benchmark suite for SystemC and TLM based Instruction Set Simulators (ISS)},'' \url{http://code.google.com/p/trap-gen}.

\bibitem{lin2004error}
S.~Lin and D.~J. Costello, \emph{{Error control coding: fundamentals and applications}}.\hskip 1em plus 0.5em minus 0.4em\relax Upper Saddle River, NJ: Pearson/Prentice Hall, 2004.

\bibitem{bossuet}
G.~Leplus, O.~Savry, and L.~Bossuet, ``{Insertion of random delay with context-aware dummy instructions generator in a RISC-V processor},'' in \emph{Proceedings of HOST}, 2022.

\bibitem{mem_frencing}
M.~Taram, A.~Venkat, and D.~Tullsen, ``{Mitigating Speculative Execution Attacks via Context-Sensitive Fencing},'' \emph{IEEE Design \& Test}, vol.~39, no.~4, pp. 49--57, 2022.

\end{thebibliography}

\end{document}